\documentclass[preprint,aps]{revtex4}
\usepackage{amssymb,amsmath}
\usepackage{graphicx,float}

\begin{document}

\title{ \Large \bf Statistical Indicators of Collective Behavior and Functional Clusters in Gene Networks of Yeast}
\author{\large  \flushleft 
Jelena \v Zivkovi\'c$^{1}$,
Bosiljka Tadi\'c$^{2}$,  
Nikolaus Wick$^{3,1}$ and 
Stefan Thurner$^{1}$
} 

\affiliation{ 
$^1$Complex Systems Research Group HNO; Medical University of Vienna;  
W\" ahringer G\" urtel, 18-20, A-1090 Vienna, Austria\\
$^2$Department  for Theoretical Physics; Jo\v{z}ef Stefan Institute; 
P.O. Box 3000; SI-1001 Ljubljana; Slovenia \\
$^3$Clinical Institute for Pathology AKH Vienna, W\" ahringer G\" urtel, 18-20, A-1090 Vienna, Austria}

\begin{abstract}
 We analyze gene expression time-series  data of yeast ({\it S. cerevisiae}) measured 
along two full cell-cycles \cite{Cho}. We quantify these data by using
 $q$-exponentials, gene
expression ranking and a temporal mean-variance analysis. 
We construct gene interaction networks based on correlation coefficients 
and study the formation of the corresponding giant components 
and minimum spanning trees. 
By coloring genes according to their cell function we find 
functional clusters in the correlation networks and functional branches 
in the associated trees. Our results suggest that a percolation point of functional clusters can be identified on these gene expression correlation networks.\\


\end{abstract}

\maketitle
\section{Introduction}
Gene regulatory networks describe the effective interactions between genes.
The activity of a gene, i.e., its current rate of being transcribed into RNA molecules,
can have effects on the activity levels of other genes, which will as a result
become up- or down- regulated. The sum of all up- and down- regulation relations
in the whole genome is the gene regulatory network.
The complete knowledge of the gene network would reveal a large portion of an
understanding of life. However, this goal is far from being achieved. With  present
DNA-chip technology  it is possible
to measure the transcription rates at a given point in time of an entire genome,
but even these technologies only allow a glimpse on the structure of the underlying network,
due to the underdeterminedness of the problem
\cite{biology_networks}. This situation got the physics
community interested, to statistically characterize the available data
and to (crudely) estimate the structure of the complex networks governing
gene dynamics. 
A step toward an identification of potential  gene 
interaction networks is to identify and quantify meaningful {\it statistical}
indicators of gene cooperative behavior, which is the main
purpose of the present work.
The idea is that fluctuations of gene expressions over time, e.g., during 
a cell-cycle, can be considered as an output of an interacting gene collective forming
a structured network. The hope is that a network structure estimate 
can be inferred from statistical properties. At least it should be possible to 
statistically characterize the types of potential candidate networks.

We consider the time-course expression data $x_i(t)$ for the genome of yeast {\it 
 S. cerevisiae} \cite{Cho}. 
We determine some statistical indicators of collective dynamical 
behavior of genes, such as the $q$-exponential fit of the cumulative distribution, a ranking distribution and a mean-variance analysis of differential gene expressions.
We construct and estimate the expression-correlation network from time increments of expression data and analyse clusters and spanning trees. We identify biological function of genes with use of yeast database \cite{MIPS}. We find that the resulting, correlation based clusters match considerably well with specific biological functions of genes in the cell.

\section{Scale-invariance in gene expression levels}

The genome-wide gene expression data in \cite{Cho} are given 
in the form of a matrix $x_i(t)$ in which every row 
represents one of $N=6406$ yeast genes and each column contains the time evolution of gene expression of that gene $i$. Gene expressions are measured at 17 time points, taken every 10 minutes which covers approximately two full cell-cycles.
We first properly normalize the gene expressions for each of the 17 measurements 
separately by dividing each gene expression value by the average value of gene expression for that corresponding column. In order to
avoid systematic trends in the time series we use {\it differential} expression data defined as $\Delta x_i(t)= x_i(t)-x_i(t-1)$ for each gene $i$. 
We determine the cumulative distribution $P(> \Delta x)$ for each time-interval separately and also all measurements (all entries in matrix). 
The results are given in Fig.\ 1 a.
This distribution can be fitted to a $q$-exponential form \cite{Tsallis}, 
\begin{equation}
P(\Delta x)=B{_q}\left[1-(1-q)\frac {\mid {\Delta x}\mid}{\mid {\Delta x_0}\mid}\right]^\frac{1}{1-q} \ ;\ \  q \neq 1 \quad ,
\label{pdf}
\end{equation}
where  $q$  represents the non-extensivity parameter. The fitted values 
of $q$ for the various time-intervals are in the range $1.52-1.63$. The average over all times yields $q=1.55$, potentially indicating a non-trivial collective behavior of genes along the cell-cycle. 
\begin{figure}[tb]
\begin{center}
\begin{tabular}{cc} 
\resizebox{17.6pc}{!}{\includegraphics{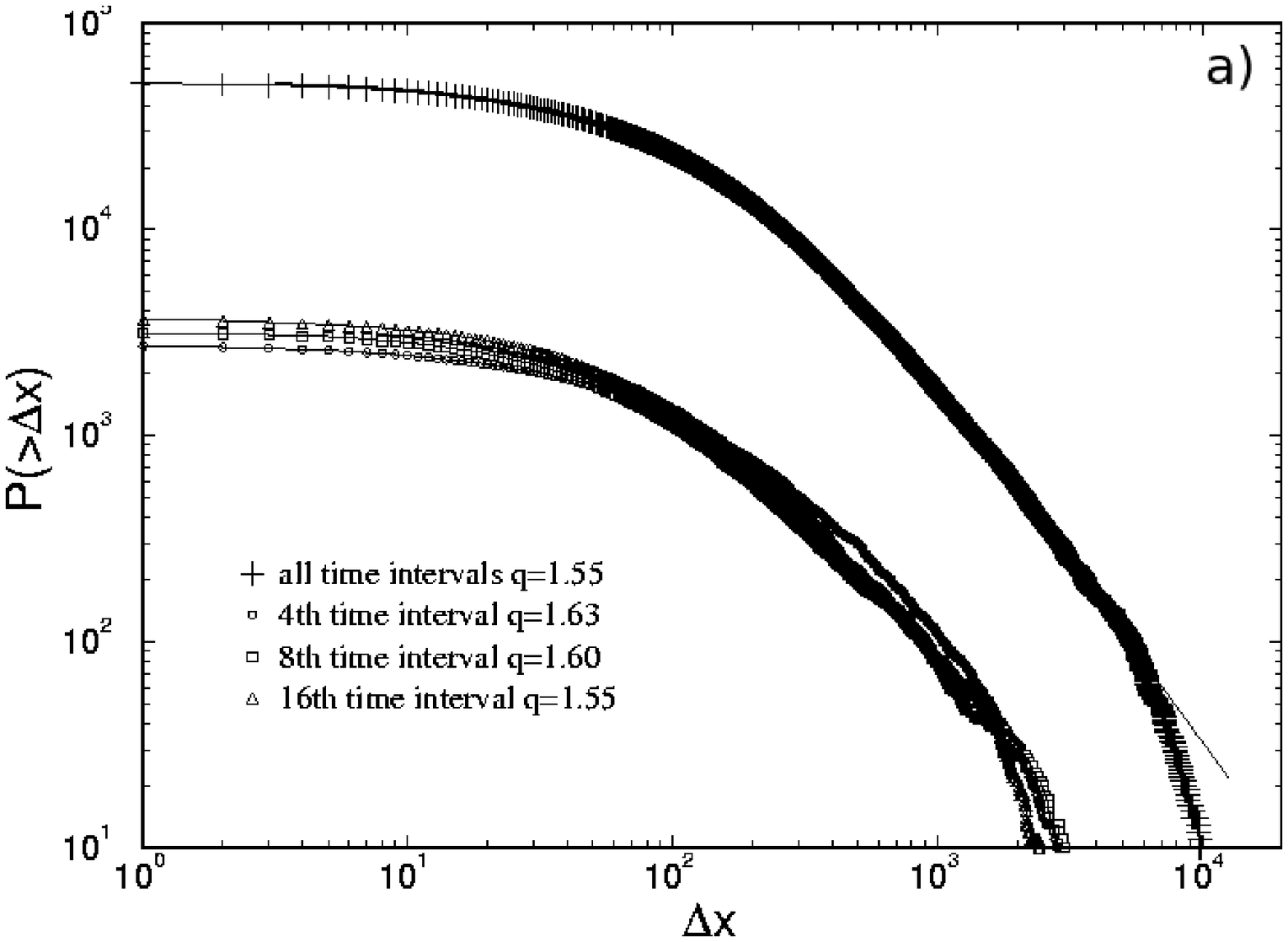}}&
\resizebox{18.4pc}{!}{\includegraphics{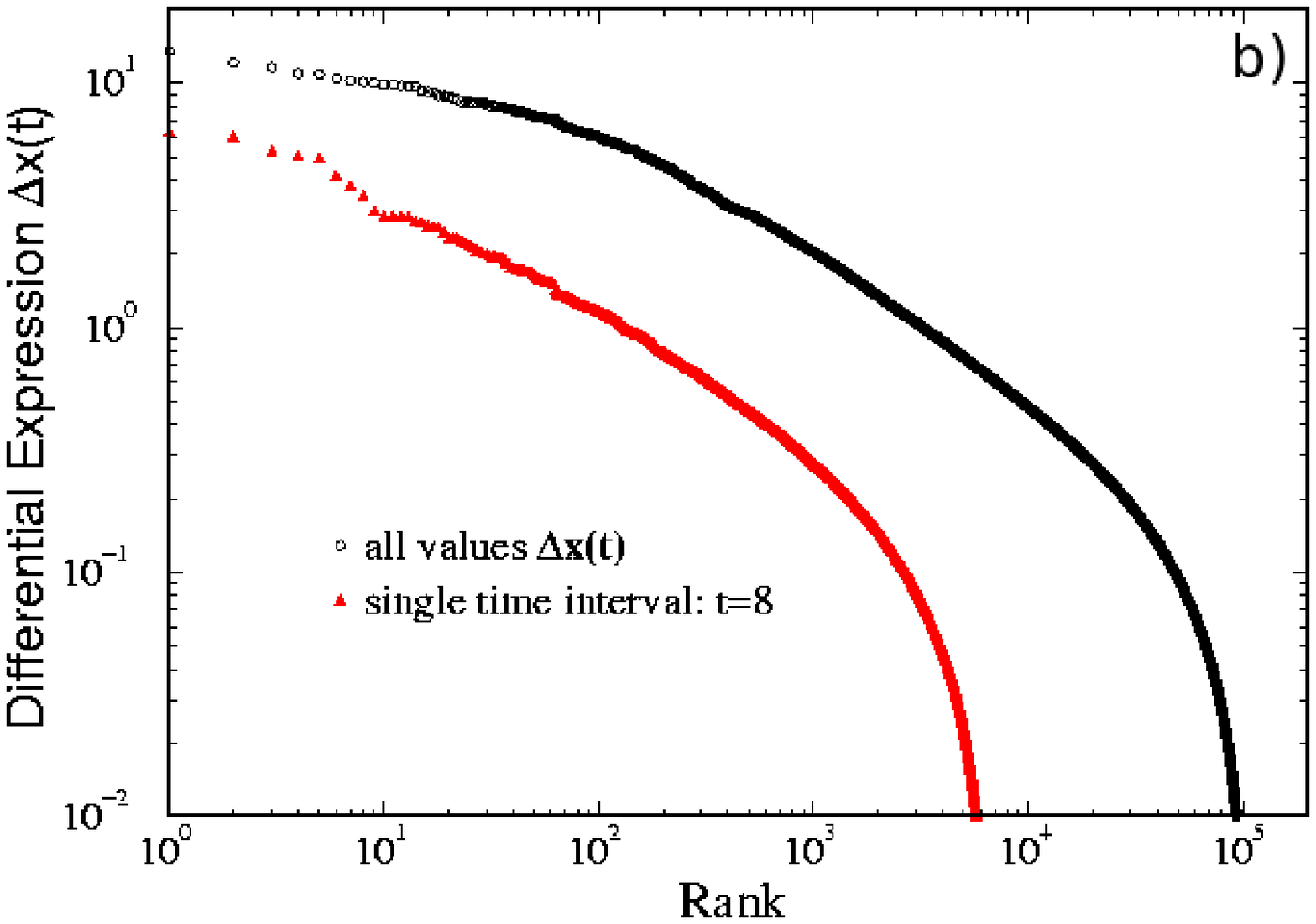}}\\
\end{tabular}
\end{center}
\caption{Histograms (a) and ranking of differential gene expressions (b). }
\label{fig-histogram}
\end{figure}
In  Fig. 1 b the ranking distribution is shown for genes according to their differential expressions at a particular time (lower curve) and for all measurements (upper curve). In both cases these curves exhibit approximate power-law regions, i.e. Zipf's law \cite{zipf}. The occurrence of Zipf's law has been found in the ranking of expression data of many other species 
\cite{Furusawa}. The results in Fig.\ 1 b indicate that the characteristic form of the distribution, in particular its slope, more or less persists  even when ranking is averaged over all-time measuremets.

\begin{figure}[tb]
\begin{center}
\begin{tabular}{c} 
\resizebox{22pc}{!}{\includegraphics{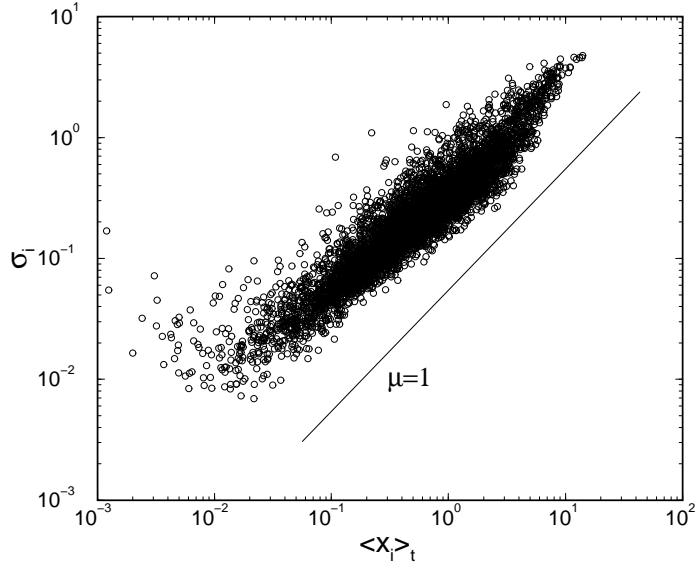}}
\end{tabular}
\end{center}
\caption{ 
Time fluctuation $\sigma_i$ plotted against time-averaged 
gene expression $\langle x_i\rangle_t$, 
for all $N$ genes.}
\label{fig-sigmah}
\end{figure}

Gene expression levels fluctuate during a cell-cycle. We calculate its
temporal mean, $\langle x_i\rangle_t = {\frac{1}{17}}\sum_t x_i(t)$ and its variance $\sigma_i ={\sqrt{{\langle {x_i}^2\rangle}_t - {{\langle {x_i}\rangle}^2}_t}}$, for all 
genes $i=1, \cdots N$. In many dynamical 
systems a relation between those quantities is found to be of the form
\begin{equation} 
 \sigma_i\sim \langle x_i\rangle ^\mu . 
 \end{equation}
In the case of driven dynamical systems on networks the scaling relation Eq.
(\ref{fig-sigmah}) holds when the values of $\mu$ depend on both, the network
topology and the driving conditions. In particular, 
many real networks seem to fall into two 'universality classes' \cite{Arg}:  
$\mu =1$, for example for scale free tree graphs and cyclic structures, and 
$\mu =1/2$, often found in weakly driven cyclic graphs. 
In Fig. 2 the temporal
variance $\sigma _i$ of the expression level $x_i(t)$ is 
plotted against its temporal mean $\langle x_i\rangle_t$ for each gene. The data yields a  slope of $\mu \sim 0.89$ which suggests a  
heterogeneous network of genes with highly driven dynamics.

\section{Gene expression networks}
\subsection{Construction of gene networks}
Measures of the statistical indicators, do not identify the network topology, however, they suggest that some collective phenomena seem to occur, which could be thought of grouped up- or down- regulations within 'clusters' of genes.

\begin{figure}[tb]
\begin{center}
\begin{tabular}{cc} 
\resizebox{18pc}{!}{\includegraphics{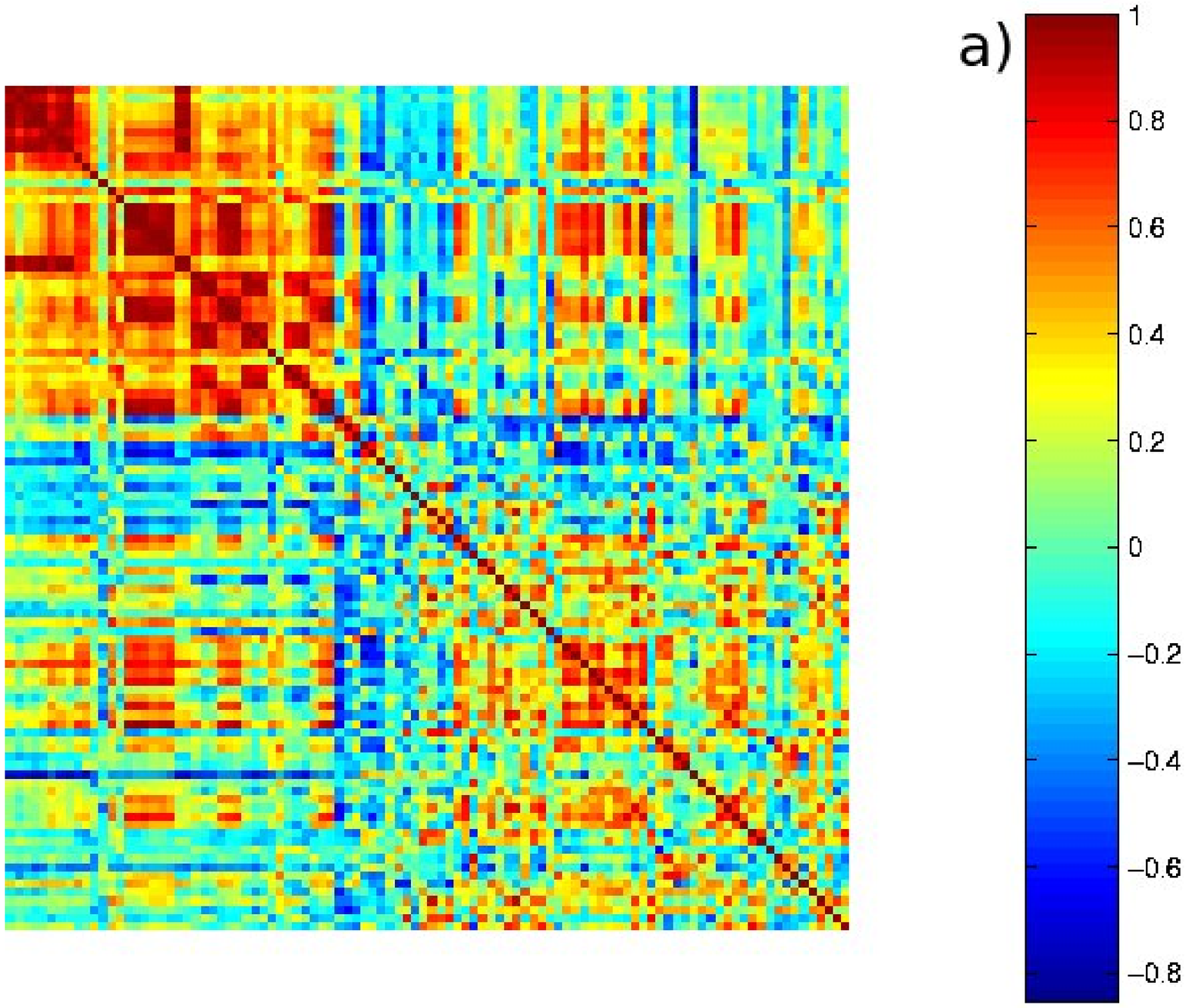}}&
\resizebox{18pc}{!}{\includegraphics{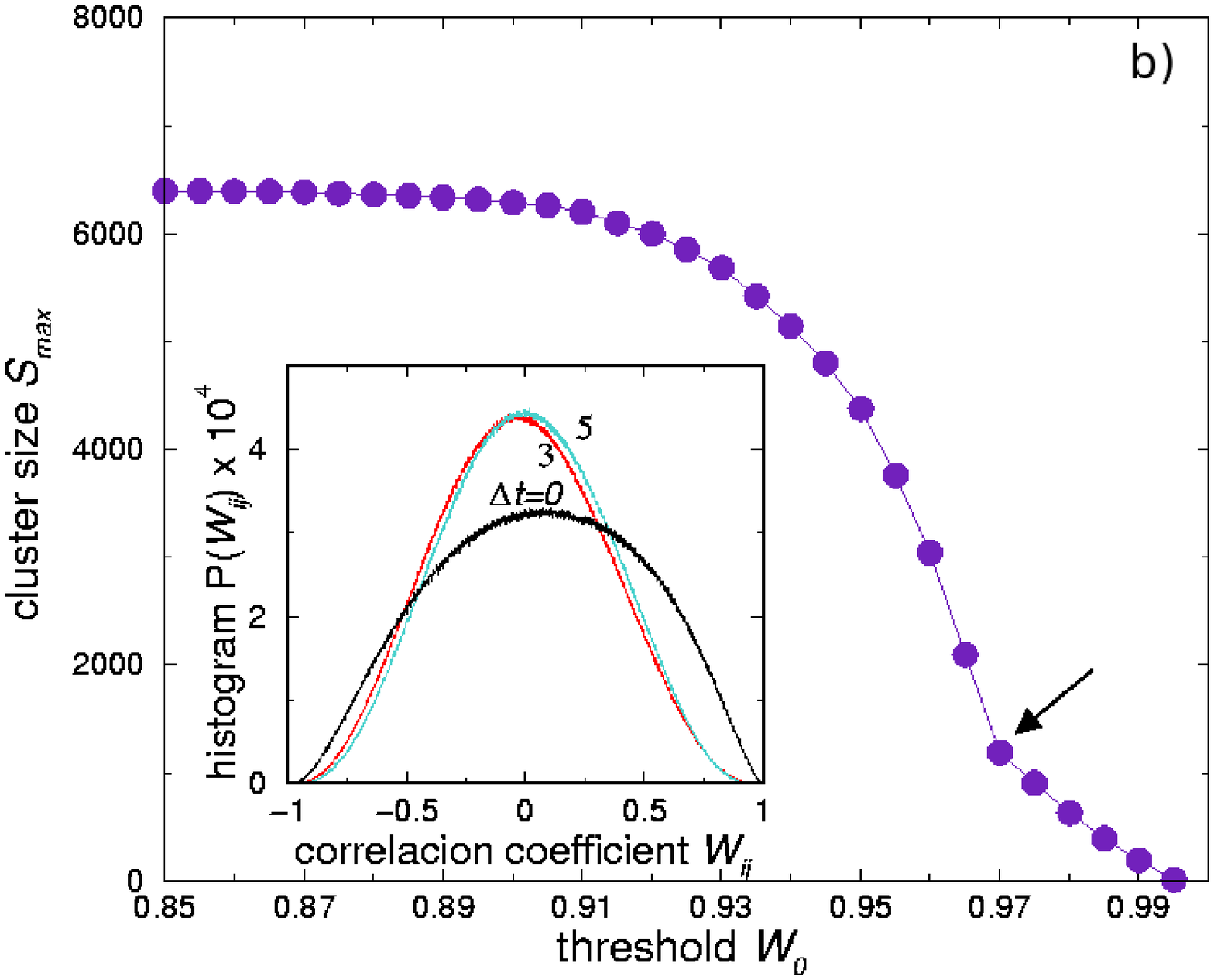}}
\end{tabular}
\end{center}
\caption{Section of the correlation matrix $W_{ij}$ (a). 
Size of the giant cluster $S_{max}$ as a function of the threshold $W_0$ 
for equal-time correlations, $\Delta t=0$, (b). Inset: histogram 
of $W_{ij}$ for different time lags, $\Delta t= 0, 3$ and $5$. }
\label{fig-sigma}
\end{figure}
As a first attempt we construct a 'gene expression network' from correlation 
coefficients of temporal differential gene expressions $\Delta x_i(t)$,  
\begin{equation}
W_{ij}(\Delta t)=\frac{\sum_{t}^{}( \Delta x_{i}(t)-\langle \Delta x_i \rangle )( \Delta x_{j}{(t+\Delta t)}-\langle \Delta x_j \rangle )}  
{
\sigma_i \sigma_j
} \quad .
\label{tau_correlations}
\end{equation}
 A section of this correlation matrix $W_{ij}$ is shown in Fig. 3 a.
The histogram of the correlation coefficients $W_{ij}$ for all pairs of 
genes are shown in 
the inset to Fig. 3 b for several time lags, $\Delta t=0, 3$ and $5$.
These distributions of correlation coefficients clearly exhibit a non-Gaussian character.
To define a network we 
select a threshold $W_0$. A link is defined to exist between genes $i$ and $j$ if their correlation exceeds the threshold, $W_{ij} > W_0$. By systematically decreasing $W_0$ we observe the formation
of a giant component, whose size $S_{max}$  is plotted against the threshold  
$W_0$ in Fig. 3 b. The conditions for the formation of the giant cluster \cite{SD_book} 
${\langle k^2 \rangle} -2{\langle k\rangle} > {\sum_{k}{k(k-2)P(k)}}>0$ 
are fulfilled at rather large values of the threshold. The size of the giant cluster increases first linearly by decreasing $W_0$, until an inflection point is reached at $W_{0}\approx 0.97$ (arrow in Fig.\ 3 b). The steep increase below this point
resembles a percolation-like behavior in which the network gradually becomes complete in the range 
$0.95 \lesssim  W_0 \lesssim 0.97$. 

\subsection{Clusters and trees}

A particular way to statistically characterize the network topology is to 
study different types of connected clusters supported by that network. 
In Figs. 4 a and c we show all clusters remaining at a thresholds of 
$W_0=0.93$ and $W_0=0.90$, respectively. A minimum cluster size of 10 nodes was chosen. Individual genes are 
nodes, colored according to their cell function \cite{MIPS}; the color map is described in the caption of Fig.\ 4.
In Figs. 4 b and d the {\it minimum spanning trees}, which are constructed from the 'distance' $d_{ij} \equiv \sqrt{ 2 (1-W_{ij})}$ are shown (see e.g \cite{mst}). The 
maximum spanning trees  computed from $W_{ij}$ directly lead to very similar 
trees (not shown), indicating that most of the dynamics is driven by 
positive correlations. 
For the threshold values in the range  $0.9\lesssim W_0\lesssim 0.97$, 
apart from the giant cluster a number of smaller clusters is present. By color-coding according to the biological functions in the cell \cite{comment}, (of which a large fraction is known for yeast \cite{MIPS}), grouping of genes into clusters occurs,
suggesting that the gene expression correlations in Eq.\ 
(\ref{tau_correlations}) captures functionally similar genes. 
By varying the threshold $W_0$ in the range between $0.9 \lesssim W_0 \lesssim 0.97$,
we detect the appearance of a color-grouping shortly below the inflection point $W_0 \approx 0.97$. Color-groupings than increases with lowering the threshold.  
For comparison, in Figs. 4 c and d we show the situation at $W_0=0.9$, where, apart from very small clusters which are removed from the figure, many new genes joined the giant component. Its minimum-distance spanning tree is also shown in Fig. 4 d.  
Genes with certain functions, in particular the 'protein synthesis' and 'cellcycle' function seem to appear in rather cohesive 
subgroups of the network. Genes with predominant 'metabolism' functions, appear more dispersed over different branches. 
All these plots are obtained for $\Delta t=0$. 
\begin{figure}[tb]
\begin{center}
\begin{tabular}{cc}
\resizebox{16pc}{!}{\includegraphics{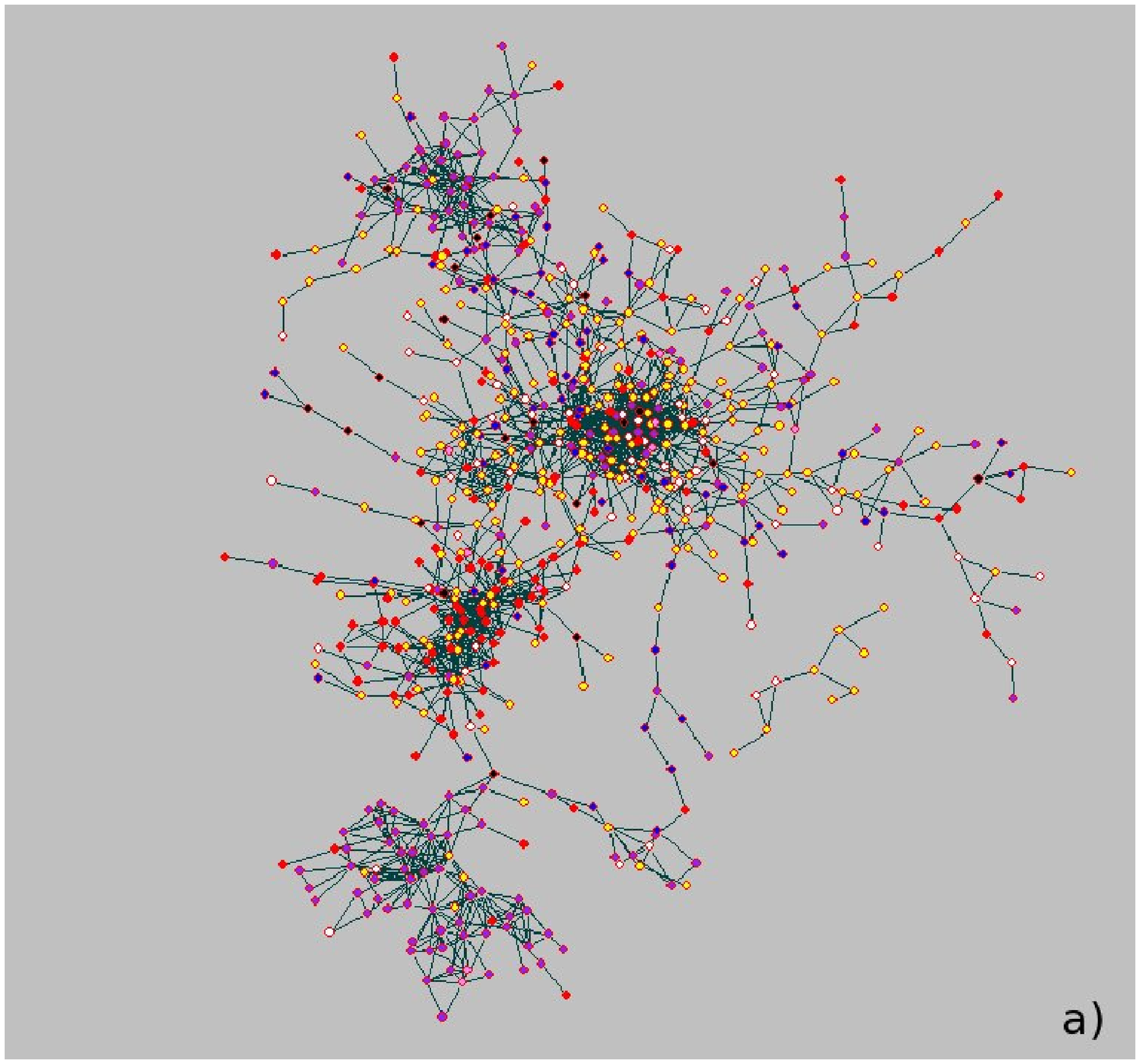}}&
\resizebox{16pc}{!}{\includegraphics{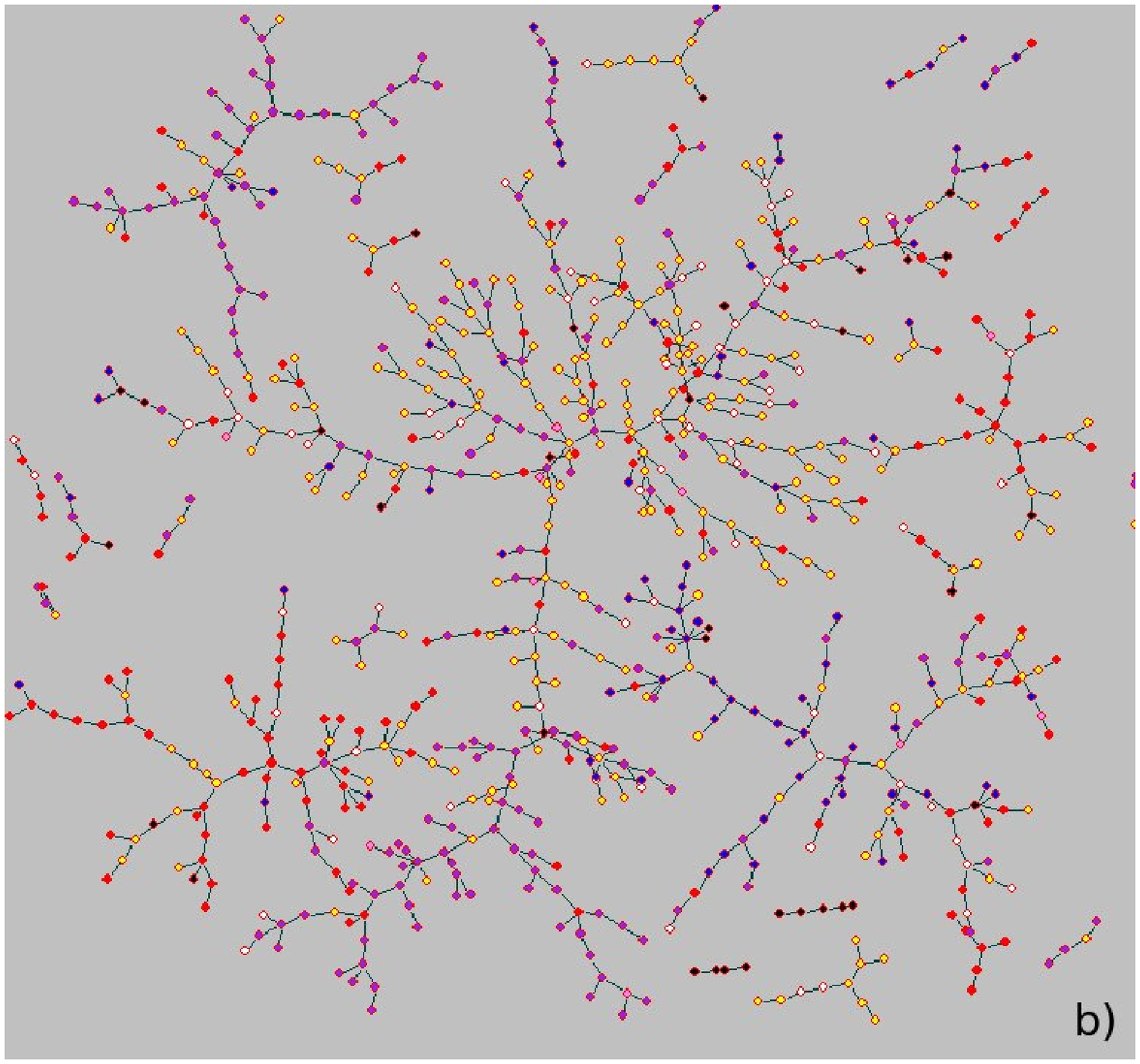}}\\
\resizebox{16pc}{!}{\includegraphics{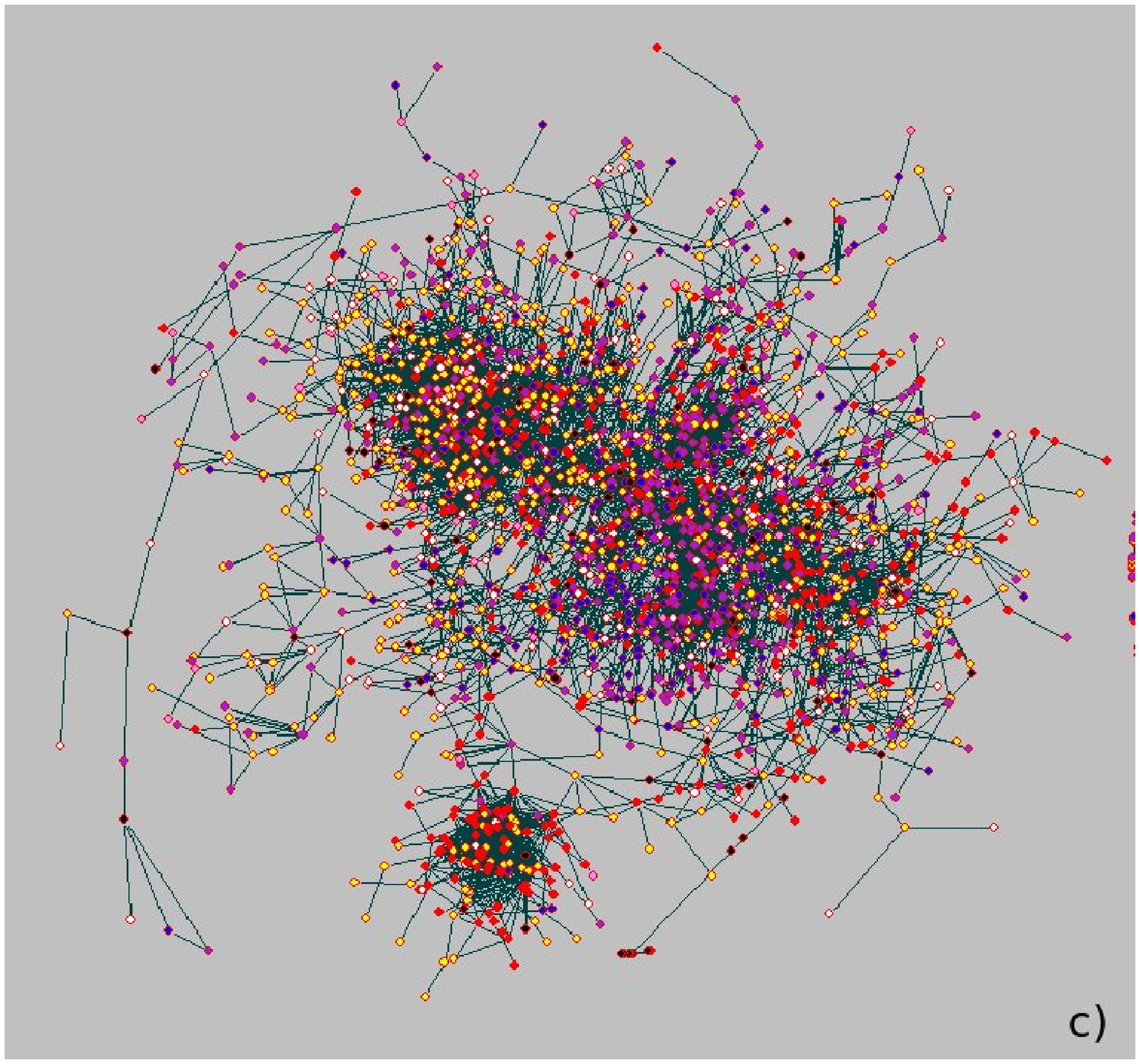}}&
\resizebox{16pc}{!}{\includegraphics{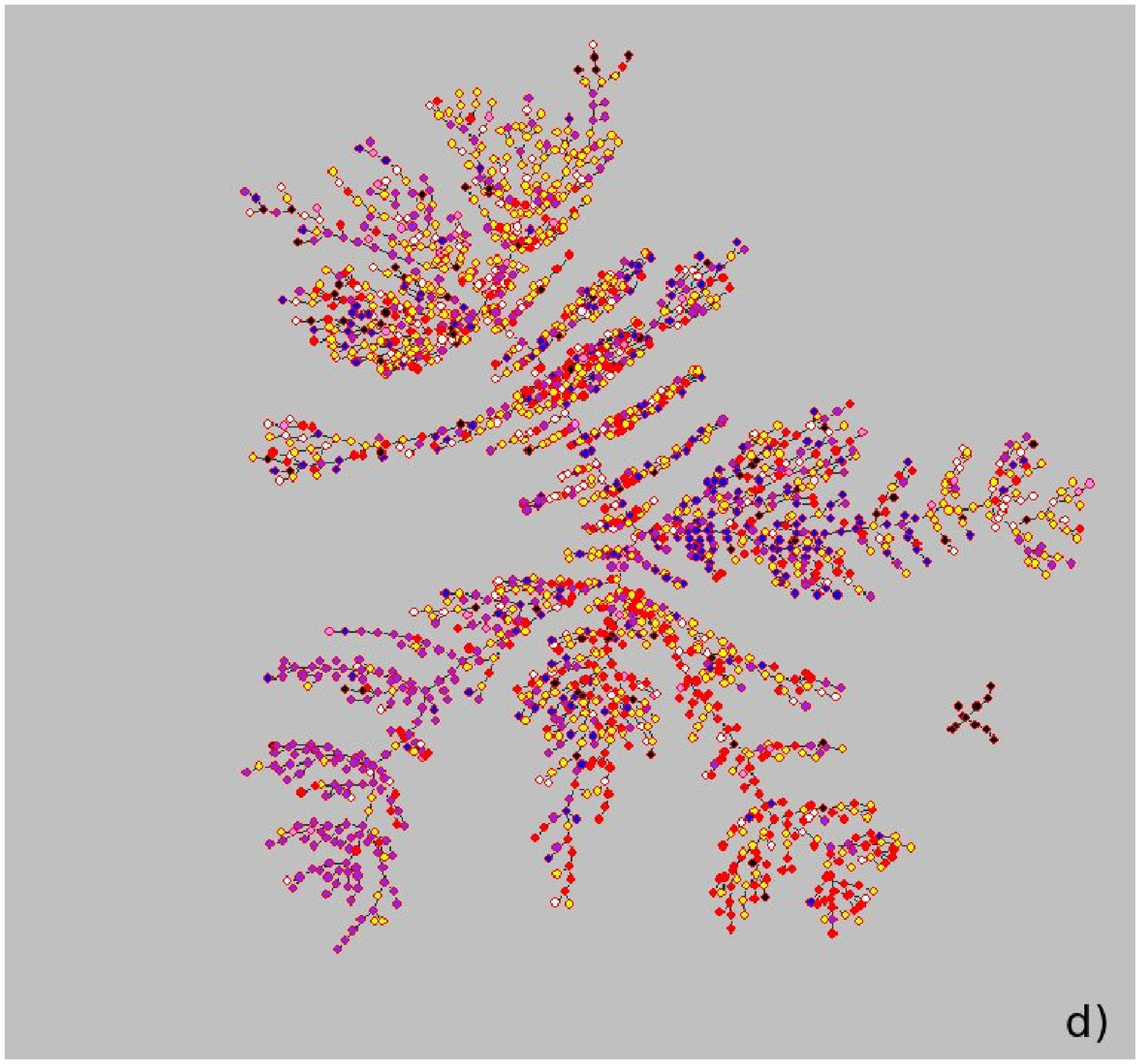}}\\
\end{tabular}
\end{center}
\caption{Gene expression networks (a and c) for $W_0=0.93$ (top) with a minimum cluster size of 10 nodes and $W_0=0.90$ (bottom). Minimum spanning trees (b and d)
for the distance measure $d_{ij}$ and same values of $W_0$. 
Genes are colored according to their  functions they fulfill in 
the cell \cite{MIPS}: yellow--metabolism; 
pink--energy household ; 
red--cellcycle; 
blue--transcription;
purple--protein synthesis 
white--cellular transport/rescue; 
black--celltype/development; 
green--unknown. 
}
\label{clusters}
\end{figure}
For $\Delta t>0$ the functional clusters and branches remain for a while 
before they gradually  disappear in the noise. This is in agreement with the 
observed character of the correlation distributions in Fig.\ 2, where 
smaller deviations from a Gaussian distribution are found for $\Delta t > 0$. 

\section{Conclusions}
In conclusion, we made several observations about the statistical nature of gene expression data which seem to suggest that at least a significant fraction of genes is up/down regulated in a highly collective manner. Indicators pointing in this direction are:
({\it i}) the cumulative distribution of differential gene expressions can be fitted to $q$-exponentials, with a non-trivial 
$q \sim 1.55$; 
({\it ii}) an approximate Zipf's law holds in the ordering distribution of differential 
expressions; 
 ({\it iii}) an almost linear mean variance dependence with  $\mu = 0.89$ signals 
tightly driven dynamics;
 ({\it iv}) the correlation matrix element distributions are non-Gaussian and non-Poisson and finally, 
 ({\it v}) even crude correlation coefficient network displays the emergence of clusters and functional branches in minimum spanning trees, which seem to be biologically relevant.

{\bf Acknowledgments:} 
B.T. acknowledges support by the Project No. P1-0044, S.T.  
by the WWTF Life Science grant LS139.
We acknowledge partial support by the bilateral project SI-AT/01/04-05.


\begin{thebibliography}{80}
\bibitem{comment}
In the available data-bases of yeast, apart from genes with unknown 
functions, often  multiple functions can be assigned to a single gene.  
To keep one color of each gene we select the first listed function in the database. We also have grouped several fuctions related to protein synthesis to a single one in order to keep the number of colors visually distinguishable.
\end{thebibliography}
\end{document}